\documentclass{article}%
\usepackage{amsmath}
\usepackage{amsfonts}
\usepackage{amssymb}
\usepackage{graphicx}%
\providecommand{\U}[1]{\protect\rule{.1in}{.1in}}

\begin{document}

\title{Einstein-Podolsky-Rosen (EPR) Correlations and Superluminal Interactions}
\author{Luiz Carlos Ryff\\\textit{Instituto de F\'{\i}sica, Universidade Federal do Rio de Janeiro,}\\\textit{Caixa Postal 68528, 21041-972 Rio de Janeiro, Brazil}\\E-mail: ryff@if.ufrj.br}
\maketitle

\begin{abstract}
The possible connection between EPR correlations and superluminal
interactions, as suggested by Bell and Bohm, is discussed using simple and
palpable arguments: (a) It is shown how an experiment based on time-like
events can allow us to answer the question \textquotedblleft Can a measurement
performed on one of the photons of an entangled pair change the state of the
other?\textquotedblright\ (b) The theorem on superluminal finite-speed causal
influences and superluminal signaling, introduced by Scarani and Gisin, is
reexamined. (c) It is shown how faster-than-light interactions and Lorentz
transformations might peacefully coexist.


\end{abstract}

\textbf{1 Introduction}

\bigskip

{\footnotesize \textquotedblleft The reason I want to go back to the idea of
an aether is because in these EPR experiments there is the suggestion that
behind the scenes some thing is going faster than light\textquotedblright. (J.
S. Bell, in The Ghost in the Atom.)}

{\footnotesize \textquotedblleft For me then this is the real problem with
quantum theory: the apparently essential conflict between any sharp
formulation and fundamental relativity\textquotedblright. (J. S. Bell, in
Speakable and Unspeakable in Quantum Mechanics. Introductory remarks at
Naples-Amalfi meeting, May 7, 1984. Reprinted in the book Speakable and
Unspeakable in Quantum Mechanics.)}

\bigskip

{\footnotesize \textquotedblleft I don't say abandon relativity theory. I'm
saying it's going to be an approximation to a much broader point of view, just
as Newtonian mechanics is an approximation to relativity\textquotedblright.
(D. Bohm, in The Ghost in the Atom.)}

{\footnotesize \textquotedblleft So long as the present type of experiment is
done, the theory of relativity will still be saved. But if we could manage to
get deeper than that then we might find that there was something going faster
than light". (D. Bohm, in The Ghost in the Atom.)}

\bigskip

Quantum entanglement, besides being an active topic of research \textrm{[1]},
with possible practical consequences in our daily life, has also raised
controversial issues related to the foundations of quantum mechanics. The term
\textquotedblleft EPR correlations\textquotedblright\textit{\ }was born from
the attempt\ by Einstein, Podolsky, and Rosen\ to demonstrate the
incompleteness of quantum mechanics\ using quantum entanglement \textrm{[2].
}Enlightening\textrm{\ }contributions to the subject have been given by
Schr\"{o}dinger \textrm{[3]} and Bohm \textrm{[4]}, and the seminal analysis
by Bell \textrm{[5]} of the EPR correlations has led to the notion of quantum
nonlocality. Physicists are divided on how to interpret this phenomenon
\textrm{[6]}, and it is not my intention to discuss the different approaches
to the theme. Here, I will start from the assumption that there are only two
reasonable explanations for the observed correlations between distant events:
a previously shared property or some kind of interaction \textrm{[7]}.
Assuming that Bell's theorem excludes the first alternative, only the second
remains, irrespective of how improbable it sounds. This immediately raises the
question about the propagation speed of this possible interaction. This
problem has been investigated by Gisin's group. Experiments have been
performed to try to determine lower bounds to this speed \textrm{[8]}, and it
has been shown that if EPR correlations result from superluminal finite-speed
causal influences then superluminal signaling is possible, at least in
principle \textrm{[9]}. In this paper I intend to show how simple and palpable
(but by no means less rigorous) reasonings help us to clarify some important
conceptual questions related to the subject. The paper is organized as
follows: In section 2 an argument supporting Bell's point of view \ according
to which EPR correlations strongly suggest that \textquotedblleft behind the
scenes something is going faster than light\textquotedblright\ \textrm{[10]}%
\ will be introduced; in section 3 the theorem on superluminal finite-speed
causal influences and superluminal signaling will be reexamined; in section 4
we will investigate how superluminal interactions and special relativity
(strictly speaking, Lorentz transformations) might peacefully coexist. (Please
note that I am not saying that EPR correlations necessarily lead to
superluminal communication)\textrm{ }Discussion and conclusion will be
presented in section 5.

\bigskip

\textbf{2 EPR Correlations and Faster-Than-Light (FTL) Interaction}

\bigskip

A question that needs to be definitively clarified is: \textquotedblleft Can a
measurement performed on one of the photons of an entangled pair change the
state of the other?\textquotedblright\ To try to answer this question\ let us
consider the ideal experiment represented in Fig.1. A source (S) emits pairs
of photons ($\nu_{1}$ and $\nu_{2}$), that propagate in opposite directions,
in the polarization-entangled state%

\[
\mid\psi\rangle=\frac{1}{\sqrt{2}}(\mid a_{\parallel}\rangle_{1}\mid
a_{\parallel}\rangle_{2}+\mid a_{\perp}\rangle_{1}\mid a_{\perp}\rangle
_{2})=\frac{1}{\sqrt{2}}(\mid b_{\parallel}\rangle_{1}\mid b_{\parallel
}\rangle_{2}+\mid b_{\perp}\rangle_{1}\mid b_{\perp}\rangle_{2})
\]%
\begin{equation}
=\frac{1}{\sqrt{2}}(\mid c_{\parallel}\rangle_{1}\mid c_{\parallel}\rangle
_{2}+\mid c_{\perp}\rangle_{1}\mid c_{\perp}\rangle_{2})=...., \tag{1}%
\end{equation}
where the ket $\mid a_{\parallel}\rangle_{1}$ ($\mid a_{\perp}\rangle_{1}$)
represents a photon $\nu_{1}$ with polarization parallel (perpendicular) to
$\mathbf{a}$, and so on \textrm{[11]}. Eq.$(1)$ emphasizes the rotational
symmetry of the situation and the fact that the photons have no privileged
polarization. A detour has been introduced to have $\nu_{1}$ always detected
before $\nu_{2}$ in all Lorentz frames. Since the detections of $\nu_{1}$ and
$\nu_{2}$ are events separated by a time-like interval, there is no doubt that
$\nu_{1}$ is \textit{really} detected first. Therefore, when $\nu_{1}$ is
found in state $\mid a_{\parallel}\rangle$ ($\mid a_{\perp}\rangle$), for
instance, we immediately know from $(1)$ that $\nu_{2}$ has been forced into
state $\mid a_{\parallel}\rangle$ ($\mid a_{\perp}\rangle$). This can be
checked using polarizer II and observing that the transmission of $\nu_{2}$
satisfies Malus' law; or, at least in principle, introducing a third photon
($\nu_{3}$) identical to $\nu_{2}$ (and in the same polarization state in
which $\nu_{1}$ has been forced, for instance) and performing a HOM experiment
\textrm{[12]}. Similarly, if $\nu_{1}$ is found in state $\mid b_{\parallel
}\rangle$ ($\mid b_{\perp}\rangle$), $\nu_{2}$ is forced into state $\mid
b_{\parallel}\rangle$ ($\mid b_{\perp}\rangle$), and so on. In other words, by
\textquotedblleft playing\textquotedblright\ with the orientation of polarizer
I it is possible to force\ $\nu_{2}$ into states $\mid a_{\parallel}\rangle$
and $\mid a_{\perp}\rangle$, or $\mid b_{\parallel}\rangle$ and $\mid
b_{\perp}\rangle$, or any other pair of mutually orthogonally polarized
states. Naturally, if the detection of $\nu_{1}$ had no influence on $\nu_{2}%
$, there would be no reason for $\nu_{2}$ to be found only in states $\mid
a_{\parallel}\rangle$ and $\mid a_{\perp}\rangle$, for instance (or only in
states $\mid b_{\parallel}\rangle$ and $\mid b_{\perp}\rangle$, and so on). On
the other hand, this influence must still be present when space-like events
are considered, since the very same correlations can be observed.%

\[%
{\parbox[b]{3.3034in}{\begin{center}
\includegraphics[
height=1.666in,
width=3.3034in
]%
{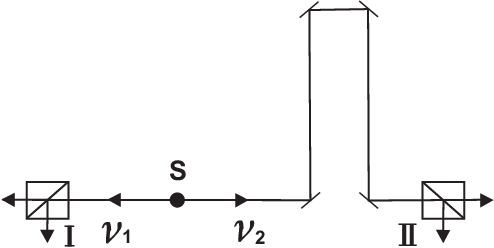}%
\\
$\textbf{Fig.1}$ {\protect\small A source (S) emits a pair of
polarization-entangled photons (}$\nu_1$ {\protect\small and }$\nu
_2${\protect\small ) that propagate in opposite directions and impinge
respectively on two-channel polarizers (I and II). A detour is introduced to
have time-like events in which }$\nu_1${\protect\small  is always detected
before }$\nu_2${\protect\small . Therefore, knowing the state in which }$%
\nu_1${\protect\small \ has been found it is possible to know the polarization
state of }$\nu_2${\protect\small  before it impinges on polarizer II. (The
arrows emerging from the polarizers indicate the positions of the detectors)}%
\end{center}}}%
\]

\bigskip

Two points are worth emphasizing in our argument. (a) It is possible to know
which photon is really detected first. This is important if we want to
investigate if, by acting on a photon of an entangled pair, we can force the
other distant photon into a well-defined state. (b) It makes it clear that the
second photon is indeed forced into a well-defined polarization state. In
fact, if $\nu_{1}$ is transmitted (reflected) at polarizer I, we know that
$\nu_{2}$ will impinge on polarizer II in state $\mid a_{\parallel}\rangle$
($\mid a_{\perp}\rangle$) (assuming that polarizer I is oriented parallel to
$\mathbf{a}$ and the entangled state is represented by $(1)$), that is, the
very same state we would obtain if $\nu_{2}$ had been transmitted (reflected)
after impinging on a polarizer oriented parallel to $\mathbf{a}$.
Therefore,\textit{ we can} \textit{predict} the outcome of an experiment
involving $\nu_{2}$ (in which we use polarizer II, for instance, or introduce
a third photon, to perform a HOM experiment). Naturally, if an observer near
polarizer II\textrm{ }had no information about $\nu_{1}$ (\textit{which is not
the case}), it would be impossible for him/her to distinguish a random
sequence of photons $\nu_{2}$\ in states $\mid a_{\parallel}\rangle$ and $\mid
a_{\perp}\rangle$ from another random sequence of photons $\nu_{2}%
$\ in\ states $\mid b_{\parallel}\rangle$ and $\mid b_{\perp}\rangle$, and so
on. [Actually, we can imagine an experiment similar to the one represented in
Fig.1, but with a very long detour and a mechanism that works in the following
way: Each time a reflected (transmitted) photon $\nu_{1}$ is detected (which
corresponds to 50\% of the detections) the first observer, near polarizer I,
can block the path of $\nu_{2}$ (\textit{because of the very long detour}). As
a consequence, the second observer, near polarizer II, will detect only
photons with linear polarization parallel (perpendicular)\ to $a$ (assuming
this is the orientation of polarizer I and the two-photon state is represented
by $(1)$). Then, it becomes evident that, whenever $\nu_{1}$ is transmitted
(reflected) $\nu_{2}$ is found in the very same state in which $\nu_{1}$ has
been forced. In other words, the second observer will receive only a sequence
of photons with linear polarization parallel (perpendicular) to $a$, other
polarizations being excluded.]

Although we have only discussed an ideal situation, quantum mechanics
predictions for EPR correlations have been corroborated by experiment
\textrm{[13]}. This strongly suggests, as advocated by Bell and Bohm, that
something is indeed going faster than light. At least, this is a possibility
that deserves to be investigated.\bigskip

\bigskip

\textbf{3 The Theorem on Superluminal Finite-Speed Causal Influences}

\bigskip

In this section, I will present an alternative and more palpable version of
the theorem on superluminal finite-speed causal influences \textrm{[9]}. The
assumption of a superluminal causal influence linking space-like separated
events raises a problem from the start: how to know which one is the cause and
which one is the effect \textrm{[14]}? A way to overcome this limitation (to
be discussed in the next section) is to assume a privileged frame in which the
\textit{actual}\ time sequence of the events can be known \textrm{[15]}. In
this frame it would be possible to determine which photon is \textit{really}%
\ detected first (even in the case of space--like events), forcing the other
(or the others) into a well-defined polarization state. Let us then consider
the experiment represented in Fig.2, which we assume as being performed in
this hypothetical coordinate system at rest in a Newtonian absolute space
\textrm{[16]}, and the three-photon polarization-entangled\ state
\textrm{[17]}%
\begin{equation}
\mid GHZ\rangle=\frac{1}{\sqrt{2}}(\mid H\rangle_{1}\mid H\rangle_{2}\mid
H\rangle_{3}+\mid V\rangle_{1}\mid V\rangle_{2}\mid V\rangle_{3}), \tag{2}%
\end{equation}
where the letters GHZ stand for Greenberger, Horne, and Zeilinger
\textrm{[18]}. Taking as a reference the plane on which the photons propagate,
the ket $\mid H\rangle_{1}$($\mid V\rangle_{1}$) represents a photon $\nu_{1}$
with horizontal (vertical) polarization, and so on. Photon $\nu_{1}$ is sent
to Alice (\textbf{A}) and $\nu_{2}$ and $\nu_{3}$ are sent to Bob (\textbf{B})
and Charlie (\textbf{C}), who work in the same lab. At instant $t_{A}$ (in the
privileged frame), \textbf{A} may decide to measure the polarization state of
$\nu_{1}$, or not; and, at instant $t_{L}>t_{A}$ (also in the privileged
frame), \textbf{B} and \textbf{C} will measure the polarization states of
$\nu_{2}$ and $\nu_{3}$. The polarizers are oriented to have the photons
emerging either in state $\mid H\rangle$ (transmitted) or in state $\mid
V\rangle$ (reflected).\ The condition $\overline{u}>l/(t_{L}-t_{A})>c$ has to
be fulfilled, where $l$ is the distance from \textbf{A }to\textbf{\ B } and
from \textbf{A }to \textbf{C, }and $\overline{u}$ is the finite superluminal
speed. (Although we are considering a very idealized situation, in which a
\textquotedblleft measurement\textquotedblright\ occurs at a precise instant
and takes no time to be completely accomplished, this will not invalidate the
essence of our reasoning.)\textrm{\ }Supposing that the correlations are
purely nonlocal, whenever \textbf{B} and \textbf{C} perform their
measurements, but \textbf{A} does not perform hers, the probability of
\textbf{B} and \textbf{C} observing the same outcome is $1/2$, since there can
be no communication between them ($\overline{u}<\infty$). On the other hand,
whenever \textbf{B} and \textbf{C} perform their measurements, and \textbf{A}
performs hers, this probability is equal to $1$, since the first measurement
forces the other two photons into the same state. Therefore, if we have in the
left lab many \textbf{A}s, and in the right distant lab the corresponding
\textbf{B}s and \textbf{C}s (and also many sources, naturally), and the
\textbf{A}s combine to take the same decision together, that is, to perform a
measurement or not, the \textbf{B}s and \textbf{C}s will know, comparing their
results (disregarding improbable statistical fluctuations), what has been
decided in the left lab before this information could reach them transmitted
by a light signal. In other words, superluminal communication would be possible.

The above discussion may give rise to two important issues. The first has to
do with the particular GHZ\ state $(2)$. It may be argued that if we had a
mixture (instead of state $(2)$)\ in which $\frac{1}{2}$ of the photons were
emitted in the state $\mid H\rangle_{1}\mid H\rangle_{2}\mid H\rangle_{3}%
$\ and $\frac{1}{2}$ in the state $\mid V\rangle_{1}\mid V\rangle_{2}\mid
V\rangle_{3}$, whenever we measured the horizontal and vertical polarizations,
we would obtain the quantum mechanical results, with no need of assuming a
nonlocal interaction. However, this would go against the quantum mechanical
formalism, which ascribe no polarization to individual photons in state
$(2)$.\ Moreover, as has been shown \textrm{[18,19]}, state $(2)$ is an
entangled state that, differently from a mixture, violates locality.
Therefore, the correlations involved in the above discussion are indeed
nonlocal \textrm{[19]}. (If we had considered a bizarre situation, in which a
fraction $p$ of the photons is emitted in state $(2)$, a fraction $\frac
{1-p}{2}$ in state $\mid H\rangle_{1}\mid H\rangle_{2}\mid H\rangle_{3}$, and
a fraction $\frac{1-p}{2}$ in state $\mid V\rangle_{1}\mid V\rangle_{2}\mid
V\rangle_{3}$, our argument would not be changed in any essential way.)%

\[%
{\parbox[b]{3.7305in}{\begin{center}
\includegraphics[
height=1.654in,
width=3.7305in
]%
{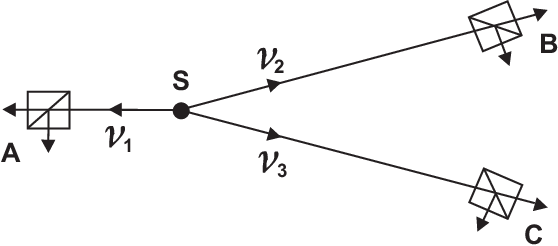}%
\\
$\textbf{Fig.2}$ {\protect\small A source (S) emits three
polarization-entangled photons (}$\nu_1${\protect\small , }$\nu
_2${\protect\small , and }$\nu_3${\protect\small ) toward three two-channel
polarizers, A (for Alice), B (for Bob), and C (for Charlie). Assuming a
superluminal finite-speed causal influence, whenever Alice decides to
\textquotedblleft remove\textquotedblright\ her polarizer, }$\nu
_2${\protect\small \ and }$\nu_3${\protect\small ,which are detected at the
same time, may be found in different polarization states. (The arrows emerging
from the polarizers indicate the positions of the detectors)}%
\end{center}}}%
\]

The second issue has to do with the duration of a measurement. In principle,
the nonlocal connection between Bob and Charlie may be arbitrarily fast. Thus,
we can imagine that the time it takes for a measurement to be accomplished is
longer than the time necessary for Bob's and Charlie's respective systems to
exchange information and to arrive, so to speak, at an agreement on which
polarization (horizontal or vertical) the photons will be found. We may even
assume that the measurement is only concluded when this agreement is reached.
In this case, it would be possible to have superluminal finite-speed causal
influences without superluminal signaling. We can even conjecture that the
superluminal speed is not constant, but depends on the distance that separates
the particles, becoming larger the larger the spatial separation between the
particles.\textrm{ }To complicate things, perfectly coincident detections
never occur. Therefore, from a strictly logical point of view it is not
immediately obvious that finite-speed superluminal interactions would
necessarily lead to the possibility of superluminal communication.\bigskip

\textbf{4 Superluminal Interaction and Breaking of the Lorentz Symmetry}

\bigskip

It might be argued that superluminal finite-speed causal influences should be
rejected from the start, since they might lead to FTL communication, and, as a
consequence, to causal paradox \textrm{[14]}. Or that they would force us to
abandon special relativity altogether. However, things may not be that simple.
In the case of the violation of parity, the reflected images of some physical
phenomena do not exist in the real world. Similarly, we may assume, \textit{in
the case of nonlocal interactions}, that the equivalence between passive and
active Lorentz transformations does not exist. In other words, some events
that occur in the moving frame cannot take place in the privileged one. More
specifically, we will assume that in the latter the speed of the
superluminal\ nonlocal interaction is a constant, irrespective of the velocity
of the source or the direction of propagation. Naturally, if Lorentz
transformations remain valid, this will not hold in the former (for instance,
this interaction will not propagate isotropically, allowing us, in principle,
to determine the velocity of the moving frame relative to the privileged
frame). Surely, the constancy of the FTL speed prevents the arising of causal
paradoxes in the privileged frame, and since events that are coincident in
space and time are still coincident under Lorentz transformations, no
paradoxes are to be expected in the moving frames; however, it can be
instructive to see how things work \textrm{[20]}. Let us consider a pair of
reference frames, $\mathbf{S}$ and $\mathbf{S}^{\prime}$, in the standard
configuration, where $\mathbf{S}$ is the privileged\textrm{\ }frame and
$\mathbf{S}^{\prime}$ moves with velocity\textrm{\ }$v<c$ along the\textrm{\ }%
$x$ axis. Assuming that the Lorentz transformations%
\begin{equation}
x^{\prime}=\gamma\left(  x-vt\right)  , \tag{6}%
\end{equation}%
\begin{equation}
t^{\prime}=\gamma\left(  t-\frac{v}{c^{2}}x\right)  , \tag{7}%
\end{equation}%
\begin{equation}
x=\gamma\left(  x^{\prime}+vt^{\prime}\right)  , \tag{8}%
\end{equation}
and%
\begin{equation}
t=\gamma\left(  t^{\prime}+\frac{v}{c^{2}}x^{\prime}\right)  , \tag{9}%
\end{equation}
connect the $\mathbf{S}$ and $\mathbf{S}^{\prime}$ coordinates
(with\textrm{\ }$\gamma=1/\sqrt{1-v^{2}/c^{2}})$, we derive the well known
result%
\begin{equation}
u^{\prime}=\frac{u-v}{1-\frac{vu}{c^{2}}} \tag{10}%
\end{equation}
and%
\begin{equation}
u=\frac{u^{\prime}+v}{1+\frac{vu^{\prime}}{c^{2}}} \tag{11}%
\end{equation}
for the velocities (to simplify the argumentation, we are only considering the
propagation along the $x$ axis).

Let us initially see how the causal paradox arises in special relativity (in
which there is no privileged frame and\textrm{\ }$\mathbf{S}$ and
$\mathbf{S}^{\prime}$ are\textrm{\ }equivalent). Let the positive
quantity\textrm{\ }$\overline{u}>c$ represent the superluminal signal speed in
$\mathbf{S}$. From $(10)$, we see that if\textrm{\ }$u=\overline{u}$, we can
choose\textrm{\ }$v$ so as to have\textrm{\ }$v\overline{u}/c^{2}>1$, which
leads to\textrm{\ }$u^{\prime}<0$\textrm{\ }(with\textrm{\ }$\left\vert
u^{\prime}\right\vert >c$ but\textrm{\ }$\neq\overline{u}$). Therefore, in
$\mathbf{S}^{\prime}$\textrm{\ }the signal propagates backwards. Similarly,
from\textrm{\ }$(11)$ we see that, if\textrm{\ }$u^{\prime}=-\overline{u}$, we
can choose a\textrm{\ }$v$ that leads to\textrm{\ }$u>0$\textrm{\ }%
(with\textrm{\ }$u>c$\textrm{\ }and\textrm{\ }$\neq\overline{u}$). That is, in
$\mathbf{S}$ the direction of propagation of the signal is reversed. It is
this change of direction when we go from $\mathbf{S}$ to\textrm{\ }%
$\mathbf{S}^{\prime}$, and then from $\mathbf{S}^{\prime}$\textrm{\ }to
$\mathbf{S}$, that is at the origin of the causal paradox. To see this, let us
consider a\textrm{\ }superluminal signal emitted from\textrm{\ }$x_{0}=0$, at
instant\textrm{\ }$t_{0}=0$, and reaching\textrm{\ }$x_{1}>0$ at
instant\textrm{\ }$t_{1}$ given by%
\begin{equation}
t_{1}=\frac{x_{1}}{\overline{u}} \tag{12}%
\end{equation}
in $\mathbf{S}$. In $\mathbf{S}^{\prime}$, the signal is transmitted
from\textrm{\ }$x_{0}^{\prime}=0$, at instant\textrm{\ }$t_{0}^{\prime}=0$,
reaching\textrm{\ }$x_{1}$ at instant%
\begin{equation}
t_{1}^{\prime}=\gamma\left(  t_{1}-\frac{v}{c^{2}}x_{1}\right)  =\gamma\left(
1-\frac{v\overline{u}}{c^{2}}\right)  \frac{x_{1}}{\overline{u}}, \tag{13}%
\end{equation}
according to\textrm{\ }$(7)$ and $(12)$. We see that\textrm{\ }$v\overline
{u}/c^{2}>1\rightarrow t_{1}^{\prime}<0$. Therefore, in $\mathbf{S}^{\prime}%
$\textrm{\ }the signal reaches\textrm{\ }$x_{1}$ before it\textrm{\ }is sent
from\textrm{\ }$x_{0}$\textrm{\ }(actually, the signal is seen to propagate
from\textrm{\ }$x_{1}$\textrm{\ }to\textrm{\ }$x_{0}$).\textrm{\ }But this
does not yet represent a paradox, since no logical contradiction is occurring.
Let us then\textrm{\ }determine the point\textrm{\ }$x_{1}^{\prime}%
$\textrm{\ }in $\mathbf{S}^{\prime}$\textrm{\ }that coincides with\textrm{\ }%
$x_{1}$\textrm{\ }at the instant at which the signal arrives. Using\textrm{\ }%
$(6)$ and\textrm{\ }$(12)$, we obtain%
\begin{equation}
x_{1}^{\prime}=\gamma\left(  x_{1}-v\frac{x_{1}}{\overline{u}}\right)
=\gamma\left(  1-\frac{v}{\overline{u}}\right)  x_{1}. \tag{14}%
\end{equation}
An observer at\textrm{\ }$x_{1}^{\prime}$\textrm{\ }can then send a return
signal with\textrm{\ }$u^{\prime}=-\overline{u}$\textrm{\ }that will take the
time of%
\begin{equation}
\delta t^{\prime}=\frac{x_{1}^{\prime}}{\overline{u}}=\gamma\left(  1-\frac
{v}{\overline{u}}\right)  \frac{x_{1}}{\overline{u}} \tag{15}%
\end{equation}
to arrive at\textrm{\ }$x_{0}^{\prime}$. This can lead to a paradox if%
\begin{equation}
t_{1}^{\prime}+\delta t^{\prime}<0, \tag{16}%
\end{equation}
that is, if the return signal reaches the origin of $\mathbf{S}^{\prime}%
$\textrm{\ }before\textrm{\ }$t_{0}^{\prime}$, namely before the first signal
has been sent. This enables an observer in this region, after receiving the
return signal, to inform another observer, at the origin of $\mathbf{S}$, not
to send the signal. As a consequence, if the signal is sent, it is possible to
send a return signal to impede the emission of the signal. That is, the signal
would be sent\textrm{\ }and not sent at the same time! Let us see the
condition\textrm{\ }$v$\textrm{\ }would have to fulfil. From\textrm{\ }$(16)$,
$(15)$,\textrm{\ }and\textrm{\ }$(13)$, we obtain%
\begin{equation}
\gamma\left(  1-\frac{v\overline{u}}{c^{2}}\right)  \frac{x_{1}}{\overline{u}%
}+\gamma\left(  1-\frac{v}{\overline{u}}\right)  \frac{x_{1}}{\overline{u}}<0,
\tag{17}%
\end{equation}
which leads to%
\begin{equation}
v>\frac{2\overline{u}}{1+\frac{\overline{u}^{2}}{c^{2}}}. \tag{18}%
\end{equation}
Since the right-hand side of\textrm{\ }$(18)$\textrm{\ }is always smaller
than\textrm{\ }$c$, it is always possible to find a\textrm{\ }$v<c$%
\textrm{\ }that satisfies the above condition; therefore, we would indeed have
a paradox.

Now let us see how the existence of a privileged frame in which the
superluminal speed is a constant does not lead to a causal paradox. Instead
of\textrm{\ }$(15)$, we have%
\begin{equation}
\delta t^{\prime}=\frac{x_{1}^{\prime}}{-\overline{u}^{\prime}}=-\gamma\left(
1-\frac{v}{\overline{u}}\right)  \frac{x_{1}}{\overline{u}^{\prime}}, \tag{19}%
\end{equation}
where the velocity of the return signal (using $(10)$)\textrm{\ }is%
\begin{equation}
\overline{u}^{\prime}=\frac{-\overline{u}-v}{1+\frac{v\overline{u}}{c^{2}}}.
\tag{20}%
\end{equation}
The condition to have a causal paradox is then%
\begin{equation}
\gamma\left(  1-\frac{v\overline{u}}{c^{2}}\right)  \frac{x_{1}}{\overline{u}%
}-\gamma\left(  1-\frac{v}{\overline{u}}\right)  \frac{x_{1}}{\overline
{u}^{\prime}}<0, \tag{21}%
\end{equation}
instead of $(17)$, where\textrm{\ }$(19)$, $(16)$, and\textrm{\ }%
$(13)$\textrm{\ }have been used. From\textrm{\ }$(20)$ and\textrm{\ }%
$(21)$\textrm{\ }we obtain%
\begin{equation}
v>c, \tag{22}%
\end{equation}
which contradicts our initial assumption that the velocity of reference frame
$\mathbf{S}^{\prime}$\textrm{\ }is slower than the velocity of light. As a
consequence, there can be no causal paradox.

The idea of a preferred frame of reference is not novel; what is essentially
new in this approach is the suggestion that superluminal interactions may
sound more palatable if we assume that, in this specific situation, there
might be no equivalence between active and passive Lorentz transformations.
(To my knowledge, this idea was discussed for the first time in Ryff, L. C.:
arXiv: 1005.5092v2 [quant-ph].)

\bigskip

\textbf{5 Discussion}

\bigskip

As we have seen, quantum mechanical formalism strongly suggests that
\textquotedblleft measuring\textquotedblright\ the polarization of a photon of
a polarization-entangled pair \textrm{[21] }forces the other into a
well-defined polarization state. This is particularly apparent when we
consider time-like events. Moreover, since the same correlations are predicted
when space-like events are taken into account, this seems to imply (as
advocated by Bell and Bohm) some kind of superluminal interaction between the
photons and, as a consequence, the existence of a privileged frame in which
the \textit{real}\ time order of the physical events could be known.
Furthermore, if the speed of this interaction is finite, then (as emphasized
by Scarani and Gisin) superluminal signaling becomes possible in principle.
Interestingly, as we have shown, assuming that the Lorentz symmetry is broken
(\textit{but Lorentz transformations remain valid}) in the case of
superluminal interactions, this does not lead to causal paradoxes or to the
abandonment of the well-established laws of physics which are Lorentz
covariant. Therefore, the possible connection between EPR correlations and
FTL\ interactions is a subject that deserves to be investigated. In my
opinion, we should keep in mind Poincar\'{e}'s wise warning: \textquotedblleft%
... le principe de relativit\'{e} physique est un fait experimental, au
m\^{e}me titre que les propri\'{e}t\'{e}s des solides naturels; comme tel, il
est susceptible d'une incessante r\'{e}vision...\textquotedblright%
\ \textrm{[22]} (the principle of physical relativity is an experimental fact,
like the properties of natural solids; as such, it is susceptible to incessant
revision) \textrm{[23]}.


\begin{thebibliography}{99}
\bibitem {21}Here, the term \textquotedblleft measuring\textquotedblright\ may
sound misleading, since, according to the quantum mechanical formalism, before
it impinges on the polarizer and is detected, no well-defined polarization can
be assigned to the photon of the entangled pair.
\end{thebibliography}
\end{document}